\documentclass{article}

\usepackage{amsmath}
\usepackage{amssymb}
\usepackage{color}
\usepackage[latin1]{inputenc}
\usepackage[T1]{fontenc}
\usepackage{authblk}

\begin{document}
\title{Comment on "A structural test for the conformal invariance of the critical 3d Ising model" by S. Meneses, S. Rychkov, J.M. Viana Parente Lopes and P. Yvernay\\
ArXiv 1802.02319}

\author[1]{B. Delamotte\thanks{delamotte@lptmc.jussieu.fr}}
\author[1]{M. Tissier\thanks{tissier@lptmc.jussieu.fr}}
\author[2]{N. Wschebor\thanks{nicws@fing.edu.uy}}
\affil[1]{Sorbonne Universit\'e, CNRS, laboratoire de physique th\'eorique de la mati\`ere condens\'ee, LPTMC, F-75005 Paris, France}
\affil[2]{Instituto de F\'isica, Facultad de Ingenier\'ia, Universidad de la Rep\'ublica, J.H.y Reissig 565, 11000 Montevideo, Uruguay}

\maketitle

\begin{abstract}
  In a recent preprint \cite{rychkov},  Meneses et al. challenge our proof that scale
  invariance implies conformal invariance for the three-dimensional
  Ising model \cite{delamotte}. We refute their arguments. We also point out a mistake in
  their one-loop calculation of the dimension of the vector operator
  $V_\mu$ of lowest dimension which is not a total derivative.
\end{abstract}

In a recent preprint, Meneses et al. \cite{rychkov} use Monte-Carlo
simulations to give evidences that conformal invariance is an emergent
symmetry in the critical domain of the three-dimensional Ising
model.\footnote{This interesting idea was already proposed in
  \cite{delamotte} but this is not acknowledged in \cite{rychkov}.}
The preprint also criticizes our proof \cite{delamotte} that scale
invariance implies conformal invariance in this
model. In the first part of this comment, we show that this criticism is actually not valid. We then point out a mistake in their 1-loop calculation. The correct result was already published in \cite{delamotte}. We rederive it in a standard 1-loop calculation as well as in the framework of Operator Product Expansion (OPE) that was used in \cite{rychkov}.

\section{Lattice correlation functions}
Our proof that scale invariance implies conformal invariance for the $d=3$ Ising model is made in two steps \cite{delamotte}. First, we show in the formalism
of the Nonperturbative Renormalization Group that scale invariance
implies conformal invariance if there is no $Z_2$-invariant local vector operator
$V_\mu(x)$ of scaling  dimension 2 (or, more generally, $d-1$ in $d$ space dimensions) which is not a total derivative. Equivalently, the implication holds if there does not exist an integrated vector operator $\int d^d x V_\mu(x)$ of scaling
dimension $-1$.\footnote{A similar sufficient condition has been proposed by Polchinski  \cite{polchinski}. Both conditions can be used equivalently in the present discussion although they do  not always coincide.} Then, using the Lebowitz
inequalities, we prove that this necessary condition is fulfilled in all dimensions for the Ising universality class by deriving a bound on the dimensions of the operators of interest. 

Our proof involves operators defined on the lattice such as
\begin{equation}
    {\cal O}(x,\{e_i\})=\phi(x+e_1)\phi(x+e_2)\phi(x+e_3)\nabla_\mu\phi(x)
\end{equation}
where $e_i$ are lattice vectors of lengths of order of the lattice spacing and
$\nabla$ is a discretized version of the gradient. In the 
"naive continuum limit" where all vectors $e_i$ are dropped, all these operators are equal to $ \phi^3(x)\nabla_\mu\phi(x)$. The authors of \cite{rychkov} pretend that we "claim that it is the naive continuum limit which determines the long-distance asymptotics" of these operators which means that we would forget the mixing of operators. This is by no means correct. First, we fully
take into account the mixing of operators in deriving our bounds, see our Equation (A1)
for instance. The mixing of composite operators ${\cal O}^{(n)}$ of degree $n$ in the field means that the long distance behavior of correlation functions involving ${\cal O}^{(n)}$  can be dominated by operators ${\cal O}^{(m)}$ with $m<n$. This is why the bound in our Equation (A1), obtained for $\vert x -y\vert\gg a$ with $a$ the lattice spacing:
\begin{equation}
     \langle \phi^\alpha(x) \phi^\beta(y)\rangle_c \le C(\alpha,\beta) G(x-y) \ \ {\rm for\  odd}\ \  \alpha, \beta 
     \label{ineq}
\end{equation}
[$G$ is the propagator and $C(\alpha,\beta)$ a constant]
is the same for all $\alpha$ and $\beta$: The operators $\phi^\alpha(x)$ with $\alpha$ odd mix with $\phi(x)$ and this is the reason why only $G$ to the power one appears in the right hand side of the inequality (\ref{ineq}). Second, what we precisely assume (see our Eq.(24) that we reproduce below for clarity) is that the correlation functions involving two operators ${\cal O}_1(x,\{e_i\})$ and ${\cal O}_2(x,\{e_i'\})$ having the same naive continuum limit have the same  -- non naive, because of mixing -- long distance behavior up to a renormalization factor:
\begin{equation}
\langle{\cal O}_1(x,\{e_i\}) O_3(y_3 )\cdots O_n(y_n )\rangle
\sim Z_O(a)\langle {\cal O}_2(x,\{e_i'\}) O_3 (y_3 ) \cdots O_n(y_n )\rangle.
\label{eq24}
\end{equation}
 We claim in \cite{delamotte} that this assumption (i) is true to all orders of perturbation theory
 and, (ii) is currently made in Monte Carlo simulations. 

Point (i) is trivial: The model is renormalizable for $d\leq4$. As a consequence of 
general renormalization theory, the two correlation functions present in Eq.(\ref{eq24}) above have each a renormalized counterpart that exists, can be computed at any order of the epsilon-expansion and is finite when the UV regulator is removed. As already mentioned above, the renormalized operators involved in the renormalized correlation functions decompose into several operators of different scaling dimensions. As a consequence, the renormalized counterpart of the correlation function in the left-hand-side of (\ref{eq24}) is dominated at long-distance by the correlation of the leading term appearing in the decomposition of ${\cal O}_1$, that is, the operator of smallest scaling dimension with which it mixes. The same holds true for the right-hand-side of (\ref{eq24}) with ${\cal O}_2$. Now, consider the lattice as a particular regularization. The two operators ${\cal O}_1(x,\{e_i\})$ and ${\cal O}_2(x,\{e_i'\})$ being two different discretizations of the same renormalized operator $ {\cal O}_R(x)$  in the continuum can be considered as two different lattice regularizations of $ {\cal O}_R(x)$. For symmetry reasons, they  mix with the same set of operators and they therefore share the same leading operator. The long-distance behavior of the correlation functions of either ${\cal O}_1$, ${\cal O}_2$ or ${\cal O}_R$ with $O_3(y_3 )\cdots O_n(y_n )$ are thus proportional with multiplicative coefficients that, in general, depend on the lattice spacing $a$. This proves Eq.(\ref{eq24}) within perturbation theory. To the best of our knowledge, Eq.~(\ref{eq24}) has not been proven nonperturbatively \cite{rivasseau}.

An exception to the proof above occurs when the mixing with the leading operator
of either ${\cal O}_1$ or ${\cal O}_2$ turns out to vanish for accidental reasons. In this case, the long-distance  behaviour of the correlation function comes from the  subleading operator which makes all the bounds we derived in \cite{delamotte} to be again satisfied.

As for Monte Carlo simulations, it is clear that if Eq. (\ref{eq24}) were not correct, two different
lattice discretizations of, say, $ \nabla_\nu \phi(x)$ would generically lead to two different long distance behaviors of the correlation functions involving this operator. This would trivially violate universality. For instance, in their Monte Carlo simulations, Meneses et al. discretize the gradient in their equation below (2.1) in the following way:
\begin{equation}
   \nabla_\nu s(x)= \frac 12 s(x+e_\nu)-  \frac 12 s(x-e_\nu)
\end{equation}
but they could have chosen as well another discretization of the gradient such as for instance
\begin{equation}
\nabla_\nu s(x)= -\frac{1}{12}s(x+2e_\nu)+\frac{2}{3}s(x+e_\nu)- \frac{2}{3}s(x-e_\nu)+\frac{1}{12}s(x-2e_\nu).
\end{equation}
Having the same naive continuum limit, these two discretizations lead at long distance to the same behavior for the correlation functions  involving this operator. This is nothing but a particular case of our Eq. (\ref{eq24}). Thus, the authors of \cite{rychkov} use -- implicitly -- the same assumption as us, which is fine.

Let us now examine what Menenes et al. consider as a counter-example to our bound, see their Eq.(B.2) and below. They consider a complex field 
$\phi(x)$ and the $O(2)$ model\footnote{or, equivalently $U(1)$ supplemented by the mirror symmetry: $\phi\leftrightarrow\phi^\star$} defined on the lattice and involving this field. They then introduce the operators
\begin{equation}
   \phi(x+e_1)\phi(x+e_2)\phi^*(x+e_3)\nabla_\mu\phi^*(x)+
    \phi^*(x+e_4)\phi^*(x+e_5)\phi(x+e_6)\nabla_\mu\phi(x)
    \label{op-O2}
\end{equation}
which  become  $\nabla_\mu[\phi(x)\phi^*(x)]^2$ in the  naive continuum limit, whatever the choice of lattice vectors $e_i$. They then write that  "if [our] argument [...] were universally valid, it would allow to conclude that any nonderivative vector operator in the O(2) model in 3d has dimension  larger than 2". They
finally exhibit the conserved current $J_\mu= \phi\partial_\mu\phi^*-\phi^*\partial_\mu\phi$ which has exactly dimension 2, and conclude that our proof is wrong. 

The problem with the argument above is the following: Our proof does not rely at all on a "universally valid" bound but, of course, on a bound for the class of integrated vector operators with specific symmetries: $Z_2$ for Ising and $O(2)$ for the model considered in \cite{rychkov}.\footnote{A generalization of our proof to vector $O(N)$ models with $N$=2, 3 and 4 will appear soon~\cite{depolsi}.} This is explicit in our Eq.(14) in \cite{delamotte}. In the O(2) case, the operator $J_\mu$ considered by the authors of \cite{rychkov} is {\it not} invariant under
$\phi\leftrightarrow  \phi^*$ and is therefore {\it not} a candidate for the class of operators that we would consider for the O$(2)$ model (it is U(1) invariant but not invariant under $\phi\leftrightarrow  \phi^*$). Of course, any lattice discretization of a O(2)-invariant operator must preserve the O(2) symmetry, a well-known fact by anyone who performs Monte-Carlo simulations. Thus, it is true that for generic vectors $e_i$, the operators in Eq.(\ref{op-O2}) mix with $J_\mu$, but this remark is pointless for our proof because we must choose the lattice vectors $e_i$ in such a way that the mirror symmetry is preserved.\footnote{One can for instance choose $e_1=e_4$, $e_2=e_5$, $e_3=e_6$.} In this case, the discretized operator (\ref{op-O2}) does not couple to $J_\mu$ and the argument in \cite{rychkov} is thus invalid.
Let us notice that in the Ising case, the situation is simpler: The symmetry is $Z_2$ and the only concern is to consider operators with an even number of fields which is, of course, what we do.

\section{Perturbative calculations}
We  finally want to make a comment on ref.\cite{rychkov}. In their Appendix B, the authors compute at one loop the smallest scaling dimension of a $Z_2$-invariant vector operator $V_\mu$ which is not a total derivative. However, their result is wrong as can be readily checked  when comparing their Eq.(1.4) with our result quoted on the fourth line of the right column of the page 012144-4 in the published version of our paper.\footnote{Meneses et al. do not quote our result in their paper.}

We first briefly sketch the calculation that led to our result \cite{delamotte} that the lowest scaling dimension of an integrated vector operator is $3+\mathcal O(\epsilon^2)$ in $d=4-\epsilon$. The idea is to add to the standard $\phi^4$ action a perturbation of the form:
\begin{equation}
\label{eq_defK}
    K_\mu\int_x \phi^3\partial_\mu \Delta \phi.
\end{equation}
This leads to a 4-point vertex of the form
\begin{equation}
\label{eq_impK}
    S^{(4)}(p_i)=6i\, K_\mu\sum_{i=1}^4(p_i)^2p_i^\mu.
\end{equation}
The calculation follows as usual. We compute the divergent part of the 1-PI 4-point vertex $\Gamma^{(4)}$ which has a contribution proportional to the momentum-dependence exhibited in Eq~(\ref{eq_impK}). At one loop, this divergence occurs in a Feynman diagram with one power of the perturbation (\ref{eq_defK}) and one power of $\phi^4$. This divergence is then absorbed in a counterterm for $K_\mu$, from which we extract the $\beta$ function:\footnote{We work here in the same conventions as Cardy \cite{Cardy:1996xt} to ease the comparison with latter calculations. In this normalization, the beta function for the $\phi^4$ coupling constant is $\beta_g=\epsilon g-72 g^2$.}
\begin{equation}
    \partial_\ell K_\mu=(-3+\epsilon)K_\mu-72 g K_\mu.
\end{equation}
Replacing $g$ by its fixed point value, we find that the 1-loop correction exactly compensates the dimensional contribution $\epsilon$. From this we deduce the result stated above for the dimension of the integrated vector operator: it remains equal to 3 at one loop.

The authors of \cite{rychkov} compute the same scaling dimension by using another approach. The main difference is that they consider {\em local} operators instead of {\em integrated} ones. In fact, as we now discuss, their result
is wrong. The origin of their mistake is that they retain only two
operators:
\begin{equation}
  \label{eq_O12}
  \begin{split}
  O_1&=\phi\partial_\mu \phi (\partial_\nu \phi)^2\\
  O_2&=\phi^2(\partial_\nu \phi) (\partial_\mu\partial_\nu \phi)    .
  \end{split}
\end{equation}
However, the operator product expansion of $O_1$ and $O_2$ with
$\phi^4$ is not closed.
Instead of the result of Appendix A of \cite{rychkov}, we find\footnote{In principle, other operators such as $\partial_\mu (\phi^2)$ should be considered
but they play no role at one loop  in dimensional regularization.}
\begin{align}
  O_1.\phi^4&=60 O_1+8 O_2+16 O_3\\
  O_2.\phi^4&=12 O_1+64 O_2+2 O_3+6 O_4
\end{align}
where:
\begin{equation}
  \label{eq_O34}
  \begin{split}
  O_3&=\phi^2\partial_\mu \phi (\partial_\nu^2 \phi)\\
  O_4&=\phi^3(\partial_\mu\partial_\nu^2 \phi)   . 
  \end{split}
\end{equation}
These OPE are complemented by:
\begin{align}
  O_3.\phi^4=36 O_3,\qquad
  O_4.\phi^4=36 O_4.
\end{align}

Since the operators $O_3$ and $O_4$ do not couple to $O_1$ and $O_2$ under
multiplication by $\phi^4$, one could expect that they do not
influence the determination of the scaling dimension of $V_\mu$. In fact, the eigenvalues 72 and 52 found in \cite{rychkov} remain eigenvalues of the problem even in presence of $O_3$ and $O_4$. The two extra eigenvalues are degenerate, with value 36.

Although the presence of $O_3$ and $O_4$ does not modify the eigenvalues, it changes their interpretation. Looking at the eigenoperators $E_\lambda=\sum_{i=1}^4\alpha_i O_i$ such that $E_\lambda . \phi^4=\lambda E_\lambda$, we find that $E_{52}$ is a total derivative:
\begin{align}
   E_{52}&= -12 O_1 +8 O_2 - {11} O_3 + 3 O_4\\
   &=\partial_\nu S_{\mu\nu}
\end{align}
where $S_{\mu\nu}$ is a traceless symmetric tensor:
\begin{align}
    S_{\mu\nu}=&\phi^3(4\partial_\mu\partial_\nu-\delta_{\mu\nu}\Delta)\phi-2\phi^2(4\partial_\mu\phi\partial_\nu\phi-\delta_{\mu\nu}\partial_\rho\phi\partial_\rho\phi).
\end{align}
Contrarily to what was stated in \cite{rychkov}, 52 is therefore not the eigenvalue we want to retain. 
The eigenoperator associated with the eigenvalue 72 is also a derivative:
\begin{align}
   E_{72}&= O_1 + O_2 + \frac{1}2 O_3 + \frac16 O_4\\
   &=\frac1{24}\partial_\mu\partial_\nu^2(\phi^4)
\end{align}
and does not either correspond to the operator $V_\mu$. The eigenoperator associated with $V_\mu$ is actually a linear combination of $O_3$ and $O_4$, associated with the eigenvalue 36. (Note that the combination $E_{36}=3O_3+O_4$ is the total derivative of the scalar $S=\phi^3\Delta \phi$).

Following \cite{Cardy:1996xt}, we readily derive that the scaling dimension of the integrated operator $V_\mu$ is $3+\mathcal O(\epsilon^2)$ while the scaling dimension of the local operator is $7-\epsilon+\mathcal O(\epsilon^2)$, in agreement with the calculation sketched at the beginning of this section.

As a consistency check, we observe that $E_{72}$, $E_{52}$ and $E_{36}$ being total derivatives, their scaling dimension should be related with the scaling dimension of their associated primary operators, which are respectively $\phi^4$, $S_{\mu\nu}$ and $S$. We have checked that, indeed, $S_{\mu\nu}.\phi^4=52 S_{\mu\nu}$ and $S.\phi^4=36 S$ and of course $\phi^4.\phi^4=72 \phi^4$, which shows that this relation indeed holds.


\begin{thebibliography}{}

\bibitem{rychkov}
S. Meneses, S. Rychkov, J. M. Viana Parente Lopes, P. Yvernay, arXiv:1802.02319.

\bibitem{delamotte}
 B.~Delamotte, M.~Tissier and N.~Wschebor,
  Phys.\ Rev.\ E {\bf 93} (2016),  012144.
  
  
 \bibitem{polchinski}  J. Polchinski, Nucl. Phys. B 303, 226 (1988).

  
  \bibitem{rivasseau}
  A. Abdesselam and V. Rivasseau, private communication.

\bibitem{depolsi} G. De Polsi, M. Tissier and N. Wschebor, in preparation.

\bibitem{Cardy:1996xt}
  J.~L.~Cardy,
  ``Scaling and renormalization in statistical physics,''
  Cambridge, UK: Univ. Pr. (1996)  (Cambridge lecture notes in physics: 3).

  
\end{thebibliography}
\end{document}